\documentclass{PoS}

\usepackage{slashed}

\newcommand{\Eq}[1]{Eq.~#1}
\newcommand{\Tab}[1]{Tab.~#1}
\newcommand{\Fig}[1]{Fig.~#1}
\newcommand{\Figs}[1]{Figs.~#1}
\newcommand{\Figure}[1]{Figure~#1}

\title{
\vspace{-100pt}
\textrm{
\small
\begin{flushright}
DESY 10-019\\
SFB/CPP-10-21
\end{flushright}\vspace{60pt}}
Status and prospects for the calculation of\\hadron structure
  from lattice QCD}

\ShortTitle{Status and prospects for the calculation of hadron
  structure from lattice QCD}

\author{\speaker{Dru B.\ Renner}\\
  NIC, DESY, Platanenallee 6, D-15738 Zeuthen, Germany\\
  E-mail: \email{dru.renner@desy.de}}

\abstract{Lattice QCD calculations of hadron structure are a valuable
  complement to many experimental programs as well as an indispensable
  tool to understand the dynamics of QCD.  I present a focused review
  of a few representative topics chosen to illustrate both the
  challenges and advances of our community: the momentum fraction,
  axial charge and charge radius of the nucleon.  I will discuss the
  current status of these calculations and speculate on the prospects
  for accurate calculations of hadron structure from lattice QCD.}

\FullConference{The XXVII International Symposium on Lattice Field Theory\\
  July 26-31, 2009\\
  Peking University, Beijing, China}

\ifpdf
\newcommand{\picangleone}{0}
\newcommand{\picangletwo}{0}
\newcommand{\picwidthone}{320pt}
\newcommand{\picwidthtwo}{300pt}
\else
\newcommand{\picangleone}{-90}
\newcommand{\picangletwo}{0}
\newcommand{\picwidthone}{252pt}
\newcommand{\picwidthtwo}{300pt}
\fi

\newcommand{\x}[1]{\langle x\rangle_{#1}}
\newcommand{\1}[1]{\langle 1\rangle_{#1}}

\newcommand{\ga}{g_{A}}
\newcommand{\gev}{\mathrm{GeV}}
\newcommand{\mev}{\mathrm{MeV}}
\newcommand{\rsq}[1]{\langle r^2\rangle_{#1}}
\newcommand{\nf}{N_\mathrm{F}}

\newcommand{\be}{\begin{equation}}
\newcommand{\ee}{\end{equation}}
\newcommand{\bd}{\begin{displaymath}}
\newcommand{\ed}{\end{displaymath}}

\begin{document}

\section{Introduction}

Our most stringent constraints on the structure of any hadron follow
from the underlying symmetries of QCD.  Translational invariance
dictates that the momentum of a hadron is the sum of the momenta of
all its constituents, giving rise to a powerful sum rule for the
nucleon.  Rotational invariance demands that the spins of hadrons have
the values $n/2$ for an integer $n$, hence the nucleon spin is
precisely $1/2$ and not $0.5$ with some experimental error.
Similarly, we know that the electric charge of the nucleon is exactly
$1$ and the net strangeness is $0$.  These statements are so
commonplace that they may even appear trivial, but each of these exact
results serves as an entry to different aspects of the dynamics of
QCD.

The momentum of the nucleon is built up from that of the quarks and
gluons.  While translational symmetry constrains the entire
contribution, the individual contributions of each quark flavor and
the gluon depend on the details of QCD.  Similarly, the spin of the
nucleon arises from the quark spin, quark orbital motion and the gluon
angular momentum, each of which are individually unconstrained by
symmetries but must sum to $1/2$.  The charge of the nucleon arises
from the quark charges that yield a total charge of $1$, but the
distribution of this charge in the spatial degrees of freedom probes
the nonperturbative structure of the nucleon.  Finally, the strange
quarks in the nucleon occur in precisely matched pairs of quarks and
anti-quarks.  Despite the net absence of strangeness, the consequences
of this hidden flavor are felt in many observables.  In the following,
I will elaborate on a few of these points and discuss what we
currently know from lattice calculations, phenomenology and
experimental measurements.  Along the way, I'll offer my opinion
regarding where our calculations may have to go to provide accurate
results for hadron structure.

The past several years have seen extensive reviews of the calculations
presented each year at the annual Lattice conference.  You can find
the few most recent reviews in~\cite{Zanotti:2008zm, Hagler:2007hu,
  Orginos:2006zz}, and a very recent and exhaustive collection of
results can be found in~\cite{Hagler:2009ni}.  Rather than duplicating
these efforts, I will instead present some of the key examples
mentioned above.\footnote{At the lattice conference I discussed
  strangeness in the nucleon.  A recent review~\cite{Young:2009ps}
  covers much of this, so I will not include these results in this
  review.}  This will unfortunately prohibit me from discussing all
the hadron structure efforts presented at Lattice 2009, but I hope
these proceedings will still provide a useful overview, nonetheless.

\section{Nucleon Momentum}

The momentum of a nucleon with energy $E$ and mass $m$ is precisely
constrained by Lorentz symmetry as $p^2 = E^2 - m^2$.  This is such a
common statement in particle physics that it seems nearly trivial to
even mention it.  However, here we want to question how this momentum
arises from the underlying QCD degrees of freedom.  In other words,
what is the distribution of this momentum among the nucleon's
constituents?  This question concerns the details of the dynamics of
QCD and presents a challenge to lattice calculations of nucleon
structure.

\subsection{Parton Distribution Functions}

The proper field theoretic response to the question above requires
constructing the momentum distributions of the nucleon's constituents,
the parton distribution functions (PDFs).  I'll provide a proper
definition of the PDFs shortly, but for the moment we want to focus on
a more intuitive understanding of the quark and gluon distributions,
$q(x)$ and $g(x)$.  In the parton model $q(x)$ or $g(x)$ would be the
probability to find a quark, of some flavor, or a gluon with momentum
$p_\mathrm{parton}=x p_\mathrm{nucleon}$.  This interpretation is
retained in QCD, however, $q(x,\mu)$ and $g(x,\mu)$ now carry a
renormalization scale $\mu$ that loosely gives the energy resolution
at which the distribution is probed.  While this muddies the picture a
bit, the PDFs, nonetheless, are well-defined universal properties of
the nucleon that are probed in many different experiments.

\subsubsection{Phenomenological Results for PDFs}

\begin{figure}[t]
\begin{center}
\includegraphics[width=\picwidthone,angle=\picangleone]{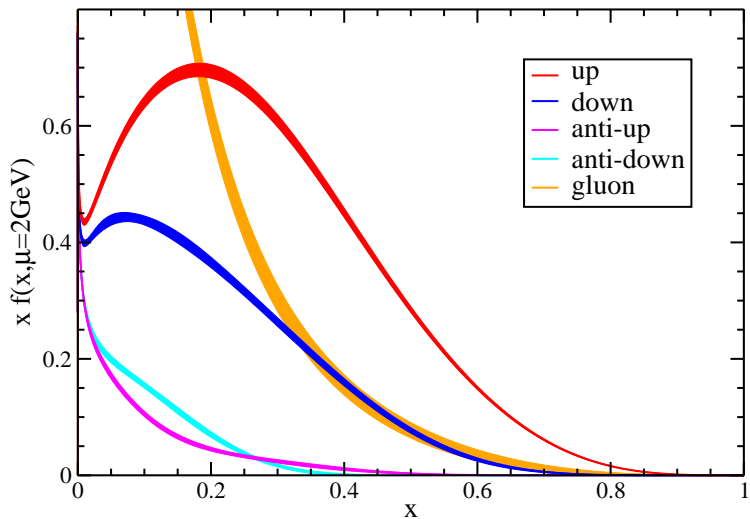}
\caption{Momentum distribution of quarks and gluons in the nucleon at
  $\mu=2~\mathrm{GeV}$.  For each parton $xf(x,\mu)$ is plotted where
  $f(x,\mu)$ is the corresponding PDF.  At large $x$, the up and down
  quarks are the largest components of the nucleon momentum.  At low
  $x$, the gluon dominates and the anti-quark distributions grow.  The
  curves were generated using the LHAPDF library~\cite{Whalley:2005nh}
  and the MSTW2008 NNLO~\cite{Martin:2009iq} dataset at a
  renormalization scale of $\mu=2~\mathrm{GeV}$.}
\label{x2dist}
\end{center}
\end{figure}

There is now a well-established industry dedicated to extracting the
PDFs from global analyses.  The analyses differ in many details,
including the order of the perturbative expansion used, the treatment
of the quark masses and the handling of the statistical and systematic
errors from the many experimental inputs.  These variations have an
impact on the precision ultimately obtained, however, these details do
not impact the discussion at hand.  Hence, as merely one example among
several, I show the PDFs from the MSTW collaboration in
\Fig{\ref{x2dist}}.

Field theory effects ultimately handicap any simple interpretation of
the PDFs, however, we can still see that the PDFs at a low scale,
$\mu=2~\mathrm{GeV}$ in \Fig{\ref{x2dist}}, retain features that one
might expect for the nucleon.  The up and down quark distributions are
the largest component for high momentum ($x\lesssim 1$).  These
distributions in fact have peaks at reasonable values of $x$, as one
might expect for a hadron dominantly composed of two up quarks and one
down quark.  However, the most prominent feature in the plot for low
$x$ is clearly the gluon distribution.  This is a clear indication of
QCD physics at work.  This has a counterpart in the simple observation
that the quark masses directly contribute only about $1\%$ of the
nucleon's mass.  Beyond the dynamics of QCD, the presence of the
anti-quarks is a striking field theory effect.  This statement may
seem mundane, but we must remember the role played by the anti-quarks
in nucleon structure to appreciate why we go through all the
difficulty to calculate hadronic matrix elements from fully dynamical
lattice QCD rather than being content with models or quenched
calculations.

\begin{figure}[t]
\begin{center}
\includegraphics[width=\picwidthone,angle=\picangleone]{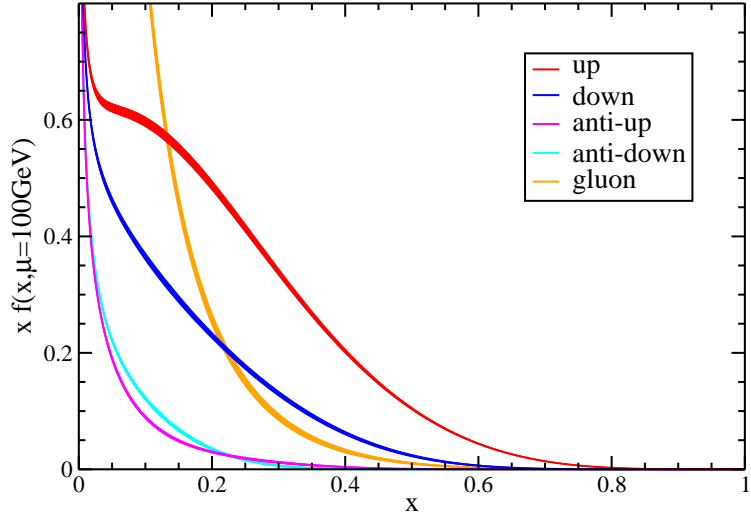}
\caption{Momentum distribution of quarks and gluons in the nucleon at
  $\mu=100~\mathrm{GeV}$.  The details are the same as in
  \Fig{\protect\ref{x2dist}} but now $\mu=100~\mathrm{GeV}$.  At high
  scales, the PDFs for the quarks and gluons mix and the simple
  non-relativistic nucleon structure, still visible at $\mu=2~\gev$,
  is suppressed at $\mu=100~\gev$.}
\label{x100dist}
\end{center}
\end{figure}

As mentioned, field theory complicates the interpretation of the PDFs.
To illustrate this, I show the same distributions again in
\Fig{\ref{x100dist}} but now evolved to a scale of
$\mu=100~\mathrm{GeV}$.  Any hint of nucleon physics is now well
hidden and all the constituents of the nucleon appear to play nearly
equally important roles.  We will return to this issue again when
discussing the evolution of the momentum fraction.

\subsubsection{Operator Definition of PDFs}

The parton distribution functions can be defined as nucleon matrix
elements of quark and gluon fields separated by light-like distances.
As an example, we record here the operator definition of the
unpolarized quark distribution $q(x,\mu^2)$~\cite{Brock:1993sz},
\be
\label{unpol}
q(x,\mu^2) = \frac{1}{2}\int\!\! \frac{d\lambda}{2\pi}\,\, e^{ix p\cdot \lambda n} 
\langle p, s | \,\, \overline{q}(-\lambda/2\,\,n)\,\, \slashed{n}\,\, 
W_n(-\lambda/2\,\,n,\lambda/2\,\,n)\,\, q(\lambda/2\,\,n)\,\, | p, s \rangle|_{\mu^2}\,.
\ee
For each flavor there are two other twist-two quark distributions, the
helicity and transversity distributions, denoted as $\Delta q
(x,\mu^2)$ and $\delta q (x,\mu^2)$ respectively.  The factor
$W_n(-\lambda/2\,\,n,\lambda/2\,\,n)$ is the Wilson line extending
along the arbitrary light-cone direction $n$ from $\lambda/2\,\,n$ to
$-\lambda/2\,\,n$,
\bd
W_n(-\lambda/2\,\,n,\lambda/2\,\,n) = 
{\cal P} exp\left( ig \int_{-\lambda/2}^{\lambda/2} d\alpha\,\, A(\alpha n)\cdot n \right)\,.
\ed
The scale, $\mu$, and scheme dependence of $q(x,\mu^2)$ comes from the
renormalization of the operator in \Eq{\ref{unpol}}.

The expression in~\Eq{\ref{unpol}} highlights the difficulty of
lattice calculations of PDFs.  Lattice QCD calculations are performed
in Euclidean space, however, the PDFs involve quark and gluon fields
separated along the light-cone.  These inherently Minkowski-space
observables are difficult to construct explicitly in Euclidean space.
But, as we will see in the next section, the moments in $x$ of these
distributions can be calculated directly in Euclidean space.

\subsection{Moments of Parton Distributions}

As just discussed, the PDFs as a function of $x$ are essentially
Minkowski-space objects.  But the moments in $x$ are related to local
operators that can be calculated in Euclidean space on the lattice.

\subsubsection{Mellin Transform:\ from $\mathbf{x}$ to $\mathbf{n}$}

The moments in $x$ are defined as follows,
\bd
\langle x^n \rangle_{q,\mu^2} = \int_{-1}^1\!\!dx\, x^n q(x,\mu^2) = \int_0^1\!\!dx\, x^n \left\{ q(x,\mu^2) - (-1)^n \overline{q}(x,\mu^2) \right\}\,.
\ed
The sign in the above equation is determined by the identification of
$q(x)$ for negative $x$ with the anti-quark distribution as
$\overline{q}(x,\mu^2) = - q(-x,\mu^2)$.\footnote{This is a frequent
  source of confusion when comparing to phenomenological
  determinations of the PDFs.  Also note that the sign is different
  for the transversity distribution but the same for the helicity
  distribution:\ $\delta \overline{q}(x,\mu^2) = \delta q(-x,\mu^2)$
  and $\Delta \overline{q}(x,\mu^2) = - \Delta q(-x,\mu^2)$.}  These
moments can be related to forward matrix elements of twist-two
operators.  First we present the complete result but then sketch the
argument in the following.
\be
\label{moments}
\langle p,s | \,\, \overline{q}(0)\,\, \gamma^{\left\{\mu_1\right.}\,\, 
iD^{\mu_2} \cdots iD^{\left.\mu_n\right\}} q(0)\,\, | p,s \rangle|_{\mu^2} 
= 2 \langle x^n \rangle_{q,\mu^2}\,\, p^{\left\{\mu_1\right.} \ldots p^{\left.\mu_n\right\}}
\ee
The brackets in $T^{\{\mu_1 \cdots \mu_n\}}$ denote symmetrization of
the indices of the tensor $T$ and subtraction of the traces.  The
precise meaning of this operation will be clarified shortly.

The derivation of \Eq{\ref{moments}} is almost elementary, but it is
so essential to the method underlying the lattice calculations that we
sketch the arguments using the unpolarized PDFs as an example.  The
first step is to introduce light-cone coordinates and gauge.  This is
a useful first step in understanding several aspects of the PDFs.  In
defining the light-cone coordinates, you introduce two light-cone
directions $n^\mu_\pm = (1,0,0,\pm 1)/\sqrt{2}$.  The two new
light-cone coordinates are given by $v\cdot n_\pm = v^{\mp}$ for a
four-vector $v$.  (Similarly, $\slashed{n}_\pm = \gamma\cdot n_\pm =
\gamma^{\mp}$.)  Light-cone gauge is the choice of $A(x)\cdot n = 0$,
which sets the Wilson line $W_n(-\lambda/2\,\,n,\lambda/2\,\,n)$ to
$1$.  Then choosing $n=n_{-}$ in \Eq{\ref{unpol}}, imposing the
light-cone gauge for $n_-$ and relabeling $\lambda$ as $y^-$ gives
\be
\label{lcpdfs}
q(x,\mu^2) = 
\frac{1}{2}\int\!\! \frac{dy^-\!\!}{2\pi}\,\, e^{ix p^+y^-} 
\langle p,s | \,\, \overline{q}(-y^-/2)\,\, \gamma^+\,\, 
q(y^-/2)\,\, | p,s \rangle|_{\mu^2}\,.
\ee

The next step in relating the moments to local operators is to use the
known limited support of $q(x)$ to the region $-1\le x\le 1$ to expand
the integration range used to define the moments,
\bd
\langle x^n \rangle_{q,\mu^2} = 
\int_{-1}^1\!\!dx\, x^n q(x,\mu^2) = 
\int_{-\infty}^\infty\!\!dx\, x^n q(x,\mu^2)\,.
\ed
The remaining steps are relatively elementary.  The key sequence
follows from combining the above with \Eq{\ref{lcpdfs}},
\bd
\int_{-\infty}^\infty\!\!dx\, x^n e^{ixp^+y^-} = 
(ip^+)^{-n} \int_{-\infty}^\infty\!\!dx\, (\partial^+)^n e^{ixp^+y^-} =
\ed
\bd
= (-1)^n (ip^+)^{-n} \int_{-\infty}^\infty\!\!dx\, e^{ixp^+y^-} (\partial^+)^n =
(p^+)^{-n} \delta(p^+y^-) (i\partial^+)^n
\ed
where $\partial^+ = \partial / \partial y^-$.  The above manipulations
must be understood as acting under the $\int dy^-$ in
\Eq{\ref{lcpdfs}}.  The final result is
\bd
\langle x^n \rangle_{q,\mu^2} = 
2^{-1} (p^+)^{-(n+1)} \langle p,s | \,\, \overline{q}(0)\,\, \gamma^+\,\, 
(i\partial^+)^n q(0)\,\, | p,s \rangle|_{\mu^2}\,.
\ed
By inspection, this expression can be seen as the light-cone
coordinate, light-cone gauge form of the following
\bd
\langle p,s | \,\, \overline{q}(0)\,\, \slashed{n}\,\, 
(in\cdot D)^n q(0)\,\, | p,s \rangle|_{\mu^2} 
= 2 \langle x^n \rangle_{q,\mu^2} (n\cdot p)^{n+1}\,.
\ed
This expression relates $\langle x^n \rangle_{q,\mu^2}$ to diagonal
matrix elements of local operators.  The more familiar form follows
from writing the above as
\bd
n_{\mu_1} n_{\mu_2} \ldots n_{\mu_n} 
\langle p,s | \,\, \overline{q}(0)\,\, \gamma^{\mu_1}\,\, 
iD^{\mu_2} \cdots iD^{\mu_n} q(0)\,\, | p,s \rangle|_{\mu^2}
= 2 \langle x^n \rangle_{q,\mu^2}\,\, 
n_{\mu_1} \ldots n_{\mu_n} p^{\mu_1} \ldots p^{\mu_n}\,.
\ed
This form makes it clear precisely what the symmetrization and trace
removal in \Eq{\ref{moments}} means.

\subsubsection{Phenomenology of $\mathbf{\x{u-d}}$}

\begin{figure}[t]
\begin{center}
\includegraphics[width=\picwidthone,angle=\picangleone]{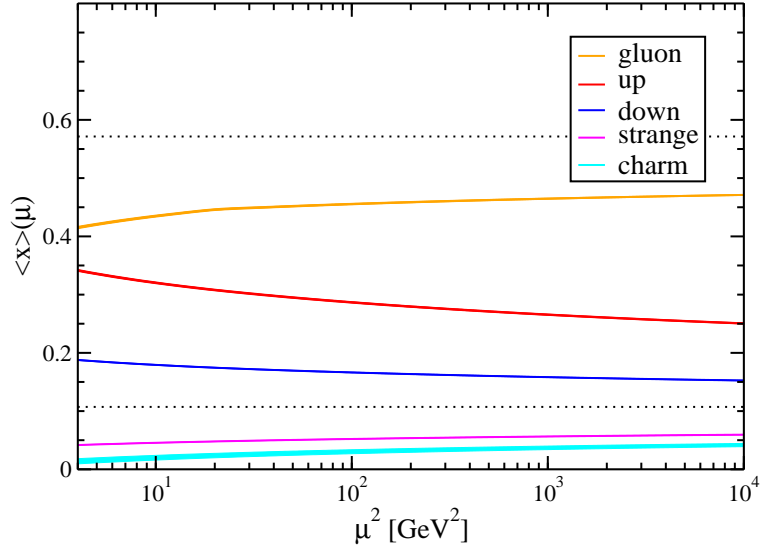}
\caption{Running of the momentum fraction of the quarks and gluons in
  the nucleon.  As $\mu$ increases, the gluon momentum fraction
  $\x{g}$ increases towards the asymptotic value of $4/7$ and $\x{q}$,
  for each of the quark flavors, approaches the common value of
  $3/28$.  (These values hold for the $4$ flavor theory.)  Despite the
  unique limiting values for $x{}$, the non-perturbative input at low
  $\mu$ clearly dictates the values of the momentum fraction for each
  parton over many orders of magnitude.  The results come from the
  LHAPDF library~\cite{Whalley:2005nh} using the
  CTEQ6.6C~\cite{Nadolsky:2008zw} dataset.}
\label{avex}
\end{center}
\end{figure}

The global analysis illustrated in \Figs{\ref{x2dist}} and
\ref{x100dist} can also be used to examine the moments of the PDFs.
This time taking as an example the results from the CTEQ
collaboration, I plot the results for $\x{q}$ and $\x{g}$ in
\Fig{\ref{avex}}.
These results follow from numerically integrating curves similar to
those in \Figs{\ref{x2dist}} and \ref{x100dist}.  Furthermore, the
state-of-the-art results for the particular quantity of interest to
us, $\x{u-d}$, are collected in \Tab{\ref{phenox}}.  The
phenomenological analyses often present results in terms of the
so-called valence distribution $q_v(x) = q(x) - \overline{q}(x)$ and
the anti-quark distribution $\overline{q}(x)$, but \Tab{\ref{phenox}}
gives the correct combination that is comparable to lattice results.
\begin{table}[t]
\begin{center}
\begin{tabular}{|l|l|l|}\hline
      & $\left[\,x\,\right]_{u_v-d_v}$        & $\x{u-d}$ \\\hline
ABMK  & $0.1790 \pm 0.0023$                & $0.1646 \pm 0.0027$ \\\hline
BBG   & $0.1747 \pm 0.0039$                & $0.1603 \pm 0.0041$ \\\hline
JR    & $0.1640 \pm 0.0060$                & $0.1496 \pm 0.0062$ \\\hline
MSTW  & $0.1645 \pm 0.0046$\footnotemark{} & $0.1501 \pm 0.0048$ \\\hline
AMP06 & $0.1820 \pm 0.0056$                & $0.1676 \pm 0.0058$ \\\hline
BBG   & $0.1754 \pm 0.0041$                & $0.1610 \pm 0.0043$ \\\hline
\end{tabular}
\caption{Phenomenological values for $\x{u-d}$ at $\mu=2~\gev$.  For
  each calculation we give the moment of the non-singlet valence
  distribution, denoted by $\left[\,x\,\right]_{u_v-d_v}$.  (The
  square brackets denote simply the integral over the region $0\le
  x\le 1$ as opposed to the region $-1\le x\le 1$ used in the angular
  brackets.)  These values were collected in~\cite{Alekhin:2009ni}.
  The original references are \cite{Alekhin:2009ni} (ABMK),
  \cite{Blumlein:2006be,Blumlein:2004ip} (BBG),
  \cite{JimenezDelgado:2008hf} (JR), \cite{Martin:2009bu} (MSTW),
  \cite{Alekhin:2006zm} (AMP06) and
  \cite{Blumlein:2006be,Blumlein:2004ip} (BBG $\mathrm{N^3LO}$). This
  is combined with the result
  $\left[\,x\,\right]_{\overline{u}-\overline{d}} = -0.0072 \pm
  0.0007$ from \cite{Alekhin:2009ni} to produce a result for
  $\x{u-d}$, which can be compared to lattice calculations.}
\label{phenox}
\end{center}
\end{table}
\footnotetext{The error given in~\cite{Martin:2009bu} is asymmetric,
  but here we take only the upper error to compare to the lattice
  results that are all higher than the phenomenological values.}

What is particularly interesting about the momentum fraction $\x{}$ is
that it obeys a sum rule,
\be
\label{momsumrule}
1 = \sum_q\, \x{q,\mu^2} + \x{g,\mu^2}
\ee
where $\sum_q$ is the sum over all relevant flavors.  This sum rule
imposes constraints on the scale evolution of $\x{\mu^2}$ such that
$\x{\mu^2}$ asymptotically approaches a perturbatively calculable
limit for large $\mu$.
\bd
\lim_{\mu\rightarrow\infty} \x{q,\mu^2} = \frac{3}{16+3\nf}
\ed
\bd
\lim_{\mu\rightarrow\infty} \x{g,\mu^2} = \frac{16}{16+3\nf}
\ed
For $\nf=4$, we have $\lim\,\x{q}=3/28\approx 10\%$ and
$\lim\,\x{g}=4/7\approx 60\%$.  However, we can clearly see in
\Fig{\ref{avex}} that the asymptotic results for $\x{}$ bear little
resemblance to the results at any reasonable value of $\mu$.  In fact,
the hierarchy shown in \Fig{\ref{avex}} is quite clear.  The gluons
carry about $40\%$ of the nucleon momentum, the up and down quarks
carry about $35\%$ and $20\%$, and the strange takes most of the
remaining $5\%$.  It is precisely this pattern that we eventually hope
to understand from lattice calculations.

\subsection{Lattice Calculation of $\mathbf{\x{u-d}}$}

The lowest non-trivial moments of the unpolarized quark PDFs are
$\x{q}$.\footnote{The lowest moments $\langle 1\rangle_q$ are known in
  terms of the quark valence structure of the nucleon.}  These are the
quark contributions that enter the momentum sum rule in
\Eq{\ref{momsumrule}}.  Here I use the particular combination
$\x{u-d}$ as a benchmark observable to evaluate the lattice
calculation of moments of PDFs.
\bd
\left.\langle p, s| \overline{q}\gamma^{\left\{\mu\right.} iD^{\left.\nu\right\}} \tau^3 q |p,s\rangle\right|_{\mu^2} = 2 \x{u-d,\mu^2} p^{\left\{\mu\right.} p^{\left.\nu\right\}}
\ed
Here $q=(u,d)$ is the doublet of the light quarks.  The flavor
combination $u-d$ eliminates disconnected diagrams, which would
otherwise require a substantial computational investment.
Additionally, this combination also eliminates any mixing with gluonic
operators, thus greatly simplifying the renormalization of this
quantity.~\footnote{Strictly speaking, the disconnected diagrams and
  the mixing with gluons only vanish for lattice actions with an exact
  flavor symmetry.}  Without the mixing, this observable then only
requires a multiplicative renormalization.  Finally, you can calculate
the bare matrix elements using nucleons with $\vec{p}=0$.  Hence,
$\x{u-d}$ is basically the most accurate scale-dependent observable in
nucleon structure that we calculate on the lattice.

\begin{figure}[t]
\begin{center}
\includegraphics[width=\picwidthtwo,angle=\picangletwo]{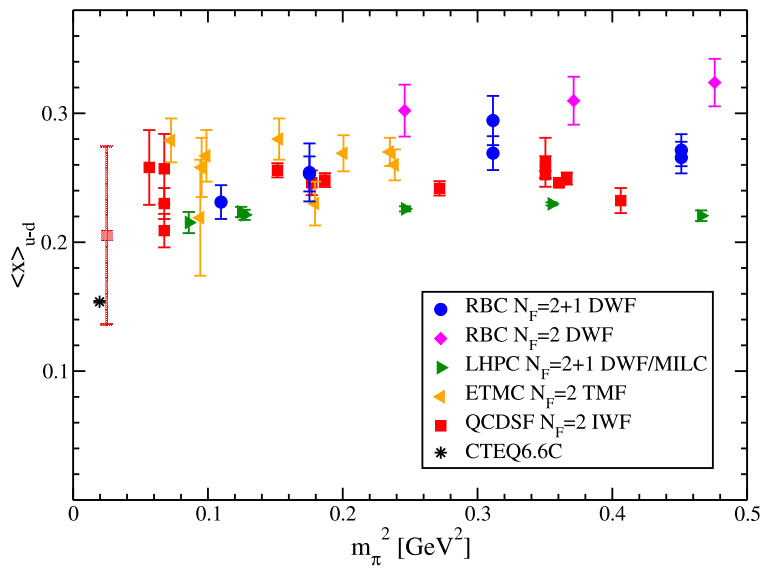}
\caption{World's dynamical lattice QCD results for $\langle
  x\rangle_{u-d}$.  The lattice calculations all overestimate the
  phenomenologically determined value.  They also show a fair amount
  of scatter amongst themselves.  The possible role of finite size and
  renormalization effects are discussed in the text.  The lattice
  results are from~\cite{Allton:2007hx,Allton:2008pn,ohta:email} (RBC
  $N_\mathrm{F}=2+1$), \cite{Lin:2008uz,Aoki:2004ht} (RBC
  $N_\mathrm{F}=2$), \cite{Hagler:2007xi} (LHPC),
  \cite{harraud:email,liu:email} (ETMC) and \cite{james:email}
  (QCDSF).  The experimental result is generated using the LHAPDF
  library~\cite{Whalley:2005nh} and the CTEQ6.6C
  dataset~\cite{Nadolsky:2008zw}.}
\label{worldx}
\end{center}
\end{figure}
\Figure{\ref{worldx}} illustrates all the dynamical lattice QCD
results for $\x{u-d}$ for pion masses less than $700~\mev$.  The
values for $\x{u-d}$, $m_\pi$ and $a$ come from a variety of already
published sources and numerous private communications.  (The
references are provided in the caption to \Fig{\ref{worldx}}.)  The
most striking feature that you observe in \Fig{\ref{worldx}} is that,
despite the community's efforts to calculate with many actions,
several lattice spacings and volumes and a broad range of pion masses
now approaching $250~\mev$, the calculations still overestimate the
experimental measurement by at least $30\%$ and maybe as much as a
factor of $2$.  The spread amongst the groups obviously suggests some
systematic variations, and I will examine two possible explanations
shortly, but it is important to note that the individual calculations
are visibly much more consistent with themselves than with the other
calculations.  Additionally, notice that some groups are beginning to
perform calculations ``at the physical point,'' as illustrated by the
QCDSF calculation in \Fig{\ref{worldx}}.  This phrase quite often
means simply $m_\pi<200~\mev$.  Nonetheless, the next generation of
lattice calculations will likely shed some much needed light on the
chiral behavior.  However, the discrepancy amongst the groups at pion
masses where we should be able to reliably calculate today is an issue
that needs to be addressed.  Without resolving these discrepancies, it
will be hard to confidently establish physical results even with
calculations approaching the physical pion mass.

Before discussing finite size and renormalization effects, I want to
make a quick comment on the lattice spacing dependence of $\x{u-d}$.
This is currently poorly studied.  Of the five calculations shown in
\Fig{\ref{worldx}}, only QCDSF has calculated beyond a single lattice
spacing.  However, even in that case, the range in $a$ that is used to
establish scaling is not large.  Their results are an encouraging hint
that lattice artifacts are not a substantial part of the discrepancy
in \Fig{\ref{worldx}}, but there is nothing universal about such
effects and all the groups must make a stronger effort to calculate at
multiple lattice spacings.

\begin{figure}[t]
\begin{center}
\includegraphics[width=\picwidthtwo,angle=\picangletwo]{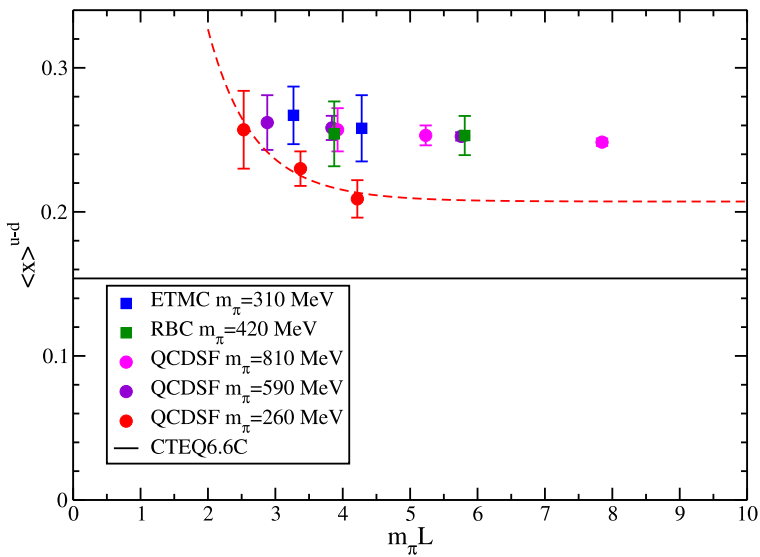}
\caption{Finite size studies for $\x{u-d}$.  The dependence of
  $\x{u-d}$ on $m_\pi L$ is shown for several calculations.  Notice
  the essentially flat behavior for $m_\pi L >4$ for all but the
  lightest pion mass.  The QCDSF calculation at the lightest pion mass
  may suggest the emergence of a more substantial finite size effect
  as $m_\pi$ is decreased.  The curve is a very simple scaling to the
  $m_\pi=260~\mev$ results and is meant solely to guide the eye.  The
  results are from the same references given in
  \Fig{\protect\ref{worldx}}.}
\label{fsx}
\end{center}
\end{figure}

A persistent concern in nucleon structure calculations is the role of
finite size effects.  In fact, the results at Lattice 2009 have added
much to this issue even if they haven't resolved it.  In
\Fig{\ref{fsx}},
I examine several finite size studies by various collaborations.
Excluding the lightest calculations at $m_\pi=260~\mev$, one observes
no statistically significant finite size effects for any of the
remaining calculations.  These results are consistent with the common
rule-of-thumb that $m_\pi L \approx 4$ is sufficient.~\footnote{My use
  of $m_\pi L$ to gauge finite size effects is, of course, not
  strictly correct.  I am loosely assuming a discussion in or near the
  chiral limit in which $1/m_\pi$ will be the dominant length scale.}
However, the recent results of QCDSF at $m_\pi=260~\mev$ potentially
stand in contrast to the finite size dependence observed at higher
pion masses.  This calculation suggests that $m_\pi L = 4$ is at best
just barely sufficient to capture the large volume limit of $\x{u-d}$.
It is possible that this observation would vanish with higher
statistics.  In fact, the three calculations at $m_\pi=260~\mev$ do
essentially agree statistically, but the trend in the results is
suggestive, especially in comparison to all the other results at
larger $m_\pi$.  Of course, it is also possible that $\x{u-d}$ would
drop even further with still larger $L$.

\begin{figure}[t]
\begin{center}
\includegraphics[width=\picwidthtwo,angle=\picangletwo]{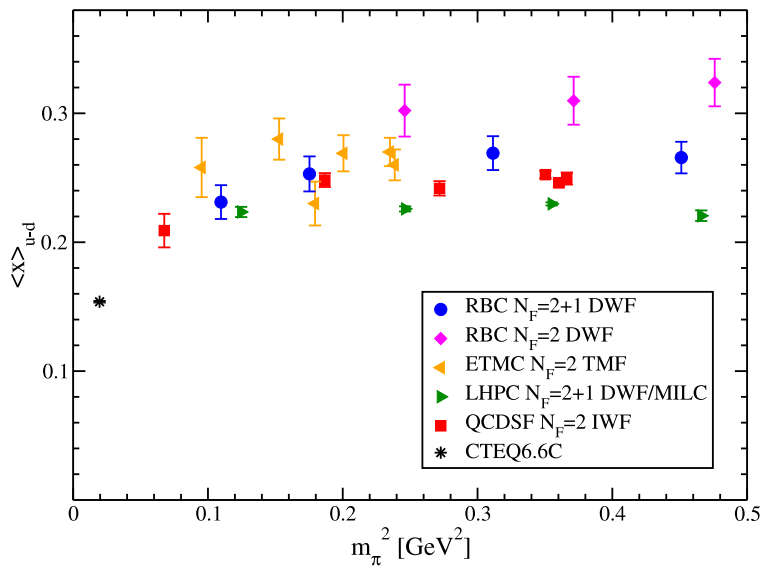}
\caption{Large volume results for $\x{u-d}$.  The results are from the
  same references given in \Fig{\protect\ref{worldx}}, but only those
  calculations with $m_\pi L>4$ are shown.  The results from each
  calculation are consistent with a smooth, nearly flat $m_\pi$
  dependence.  However there is a systematic variation between the
  calculations.}
\label{largevolx}
\end{center}
\end{figure}

In a review of this nature it is difficult to perform a detailed
infinite volume limit of all the results presented at the conference.
However, we can illustrate the impact of finite size effects by simply
making the crude restriction to $m_\pi L > 4$.  This is illustrated in
\Fig{\ref{largevolx}}.
The picture is certainly clearer and one may even be left with the
impression of some downward curvature for some of the lattice
calculations, but we must guard against wishful thinking.  Excluding
the QCDSF results, each of the groups is statistically consistent with
a linear dependence on $m_\pi^2$ with $\chi^2/\mathrm{dof} < 0.75$.
The one exception is QCDSF.  A linear fit gives $\chi^2/\mathrm{dof} =
2.1$.  This is not overwhelming evidence of non-linearity and all the
apparent curvature comes solely from the lightest point at
$m_\pi=260~\mev$.  This point is $2.8\sigma$ less than the second
lightest QCDSF point.  Increased statistics in these calculations and
confirmation of this new finite size behavior by other groups will be
necessary to understand what is happening at $m_\pi= 260~\mev$.

Given the results in \Figs{\ref{fsx}} and \ref{largevolx}, there is no
concrete conclusion that we can draw yet regarding finite size
effects, but I hope that \Fig{\ref{fsx}} will stand as a warning that
the finite size effects can, and likely do, change substantially as we
decrease the pion mass and that collaborations pushing towards the
physical pion mass must keep these effects in mind.  At this point one
may argue that we should investigate the finite size effects as
indicated from chiral perturbation theory.  It is my personal opinion
that, given an absence of any real evidence for curvature in $\x{u-d}$
and given that such non-analytic behavior is the hallmark of chiral
perturbation theory, it is theoretically questionable to force a fit
to chiral perturbation theory.  While this is my opinion, it was clear
at the lattice conference that many presenters also simply chose to
not show chiral fits.  Of course, calculations at still lighter pion
masses may show the expected chiral behavior and systematic errors at
the current pion masses may obscure this behavior, but we should keep
in mind that in the end calculations down to the physical point may be
necessary to convincingly establish results for $\x{u-d}$.

\begin{figure}[t]
\begin{center}
\includegraphics[width=\picwidthtwo,angle=\picangletwo]{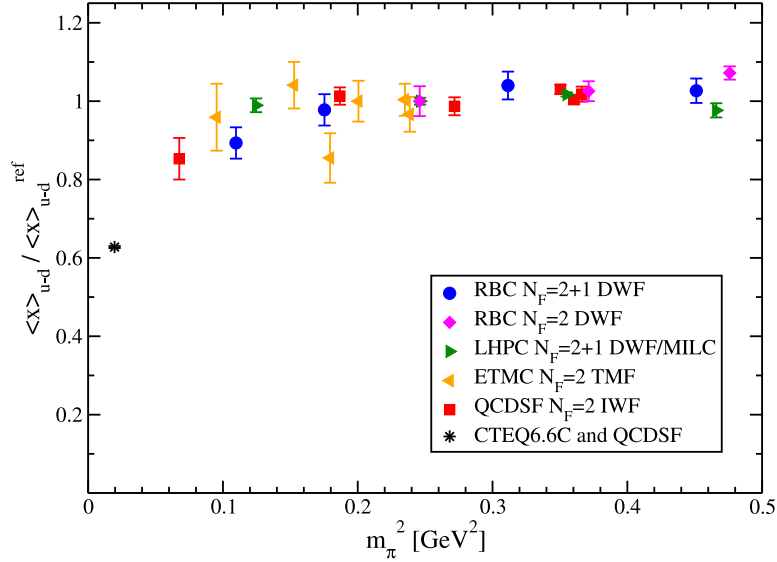}
\caption{Renormalization free ratio:\ $\langle x\rangle_{u-d} /
  \langle x\rangle_{u-d}^\mathrm{ref}$.  The calculations with $m_\pi
  L>4$ from \Fig{\protect\ref{largevolx}} are used to form the ratio
  $\langle x\rangle_{u-d} / \langle x\rangle_{u-d}^\mathrm{ref}$,
  where the reference scale is chosen as $m_\pi^\mathrm{ref} =
  500~\mev$.  This ratio eliminates the renormalization factors from
  each calculation.  The calculations are now all consistent with each
  other.  This demonstrates that a systematic error, possibly in the
  renormalization, could possibly account for the discrepancy among
  the collaborations.}
\label{ratiox}
\end{center}
\end{figure}

One obvious feature of \Fig{\ref{largevolx}} is that the results from
each collaboration have a flat behavior but there are clear shifts
between the groups.  One natural concern in this respect is the
renormalization of the operator used to determine $\x{u-d}$.  This is
multiplicative and, importantly, quark mass independent.  However it
does depend on the lattice action and hence will vary between the
different calculations.  By taking the ratio of $\x{u-d}$ to the value
$\x{u-d}^\mathrm{ref}$ at some canonical reference mass, which I
arbitrarily choose to be $m_\pi=500~\mev$, we can eliminate the
renormalization factor consistently for each calculation.  The result
of this is shown in \Fig{\ref{ratiox}},
and we find that the various calculations appear to collapse onto a
universal curve.  As with the finite size effects, we can not make a
specific conclusion, but this result is suggestive of potential
problems in the renormalization of $\x{u-d}$.  There are other
possible explanations of the systematic shifts between the groups,
such as the lattice spacing effects mentioned earlier or systematics
due to plateaus that are too short in the bare matrix element
calculations~\cite{Lin:2008uz,Yamazaki:2009zq}.  Independent of the
ultimate explanation for these problems, I hope that the difference
between \Fig{\ref{largevolx}} and \Fig{\ref{ratiox}} encourages us to
examine the systematics of our calculations.

\section{Nucleon Spin}

As discussed in the introduction, the nucleon spin is exactly $1/2$
due to rotational symmetry, and similar to the momentum fraction, it
obeys a sum rule~\cite{Ji:1998pc},
\be
\frac{1}{2} = 
\frac{1}{2} \sum_q\, \1{\Delta q,\mu^2} + \sum_q\, L_{q,\mu^2} + J_{g,\mu^2}\,,
\ee
which relates the nucleon spin to the contributions from quark
helicity $\1{q}$ and orbital angular momentum $L_q$ and a net
contribution from the gluons $J_g$.  The asymptotic evolution of the
total quark contribution $J_q=\frac{1}{2} \sum_q\, \1{\Delta q} +
\sum_q\, L_q$ and the total gluon contribution $J_g$ is given by
\bd
\lim_{\mu\rightarrow\infty} J_{q,\mu^2} = \frac{1}{2}\frac{3\nf}{16+3\nf}
\ed
\bd
\lim_{\mu\rightarrow\infty} J_{g,\mu^2} = \frac{16}{16+3\nf}\,.
\ed
For $\nf=4$, we find $\lim\, J_q = 2/7\approx 0.60\cdot1/2$ and
$\lim\, J_g = 3/14\approx 0.40\cdot1/2$.  Again we find the gluons
playing a substantial role, now in the nucleon spin and earlier in the
nucleon momentum. This is a particularly surprising conclusion in
light of the naive quark model result $\sum_q\, \1{\Delta q}=1$ and
$J_g=0$.  However, similar to the asymptotic results for the momentum
fraction, we should be suspicious that non-perturbative QCD dynamics
can alter the picture at low scales.

Unfortunately, much less is known experimentally regarding the
decomposition of the nucleon spin.  The one quantity that is known
well is the axial coupling, $\ga=\1{\Delta u}-\1{\Delta d}$, that is
measured accurately in neutron beta decay.  A recent
review~\cite{Abele:2008zz} gives $\ga=1.2750(9)$ and the most recent
PDG~\cite{Amsler:2008zzb} world average is $\ga=1.2694(28)$.  The
remaining quark contributions are essentially the only other known
pieces.  The experimental result from HERMES in
2007~\cite{Airapetian:2007mh} gives $\1{\Delta u} = 0.842(12)$,
$\1{\Delta d} = -0.427 (12)$ and $\1{\Delta s} = -0.085 (17)$.  QCD
sum rules~\cite{Balitsky:1997rs} or model
estimates~\cite{Barone:1998dx} indicate that $J_g\approx 0.25$ at low
scales.  Taking the experimental results for $\1{\Delta q}$ and this
estimate for $J_g$, we can estimate that $50\%$ of the nucleon spin is
given by the gluons with the remaining divided into $33\%$ from quark
spin and $18\%$ from quark orbital motion.  This picture is much less
certain than the momentum sum rule, but again, it provides a challenge
to the lattice QCD effort.

\subsection{Lattice Calculation of $\mathbf{\ga}$}

The nucleon axial charge can be defined by
\bd
\langle p, s | \overline{q} \gamma_\mu \gamma_5 \tau_3 q | p , s \rangle =
2 \ga s_\mu
\ed
or $\ga=\1{\Delta u-\Delta d}$.  The moments $\1{\Delta q}$ are the
lowest moments of the polarized PDFs.  For the unpolarized
distributions, the moments $\1{q}$ are fixed by the valence structure
of the nucleon, but this is not the case for the polarized
distribution.  This difference actually makes the moments $\1{\Delta
  q}$ even simpler than the momentum fractions, $\x{q}$, discussed in
the previous section.  In fact, using a lattice action with chiral
symmetry would eliminate the need to renormalize the lattice operator
used to determine $\1{\Delta q}$ altogether.\footnote{This is only
  true for the non-singlet moments, such as $\1{\Delta u-\Delta d}$.
  The axial anomaly generates mixing for the singlet moments.}
However, many lattice actions in use today do require this
renormalization, but it is not scale dependent and is in many ways
simpler than the renormalization required for $\x{q}$.  Additionally,
since the power of $x$ in $\1{\Delta q}$ is one lower than in $\x{q}$,
the lattice operator does not contain a derivative and is a simple
quark bilinear.  Consequently, the bare matrix elements can be
calculated more accurately for $\1{\Delta q}$ than for $\x{q}$.  As in
the case of $\x{u-d}$, we focus on the $u-d$ combination again to
eliminate the computationally demanding disconnected diagrams.
Furthermore, this combination gives the axial coupling $\ga=\1{\Delta
  u - \Delta d}$ that is very accurately measured in neutron beta decay.

\begin{figure}[t]
\begin{center}
\includegraphics[width=\picwidthtwo,angle=\picangletwo]{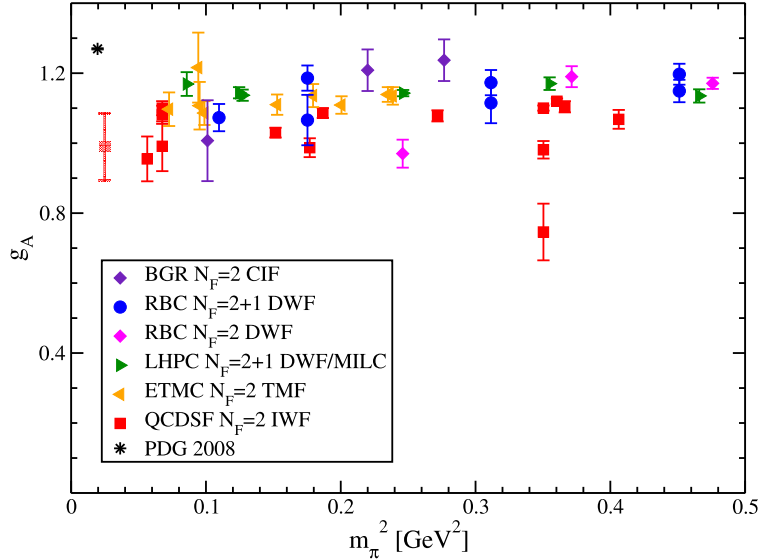}
\caption{World's dynamical results for $g_A$.  The lattice results are
  from~\cite{Gattringer:2008vj,mohler:email} (BGR),
  \cite{Yamazaki:2009zq} (RBC $N_\mathrm{F}=2+1$),
  \cite{Lin:2008uz,Aoki:2004ht} (RBC $N_\mathrm{F}=2$),
  \cite{negele:email} (LHPC), \cite{korzec:email} (ETMC) and
  \cite{james:email} (QCDSF).  The experimental result is the PDG 2008
  value~\cite{Amsler:2008zzb}.  The discrepancy between the lattice
  calculations and the experimental measurement and the scatter among
  the lattice calculations are both smaller than for $\x{u-d}$ shown
  in \Fig{\protect\ref{worldx}}.  }
\label{worldga}
\end{center}
\end{figure}

In \Fig{\ref{worldga}},
I collect all the full QCD lattice results for $\ga$, again with
$m_\pi < 700~\mev$.  Similar to $\x{u-d}$, $\ga$ serves as a benchmark
observable for lattice calculations of nucleon structure.  As in
\Fig{\ref{worldx}} for $\x{u-d}$, we find that the lattice results
show a degree of scatter and all consistently underestimate the
experimental measurement.  However, the discrepancy, between $10\%$
and $20\%$, is more mild and the scatter in the lattice results is
less severe.  In fact, as I try to argue next, it appears that this
scatter may be almost entirely accounted for by finite size effects,
at least to the current level of statistical precision.

\begin{figure}[t]
\begin{center}
\includegraphics[width=\picwidthtwo,angle=\picangletwo]{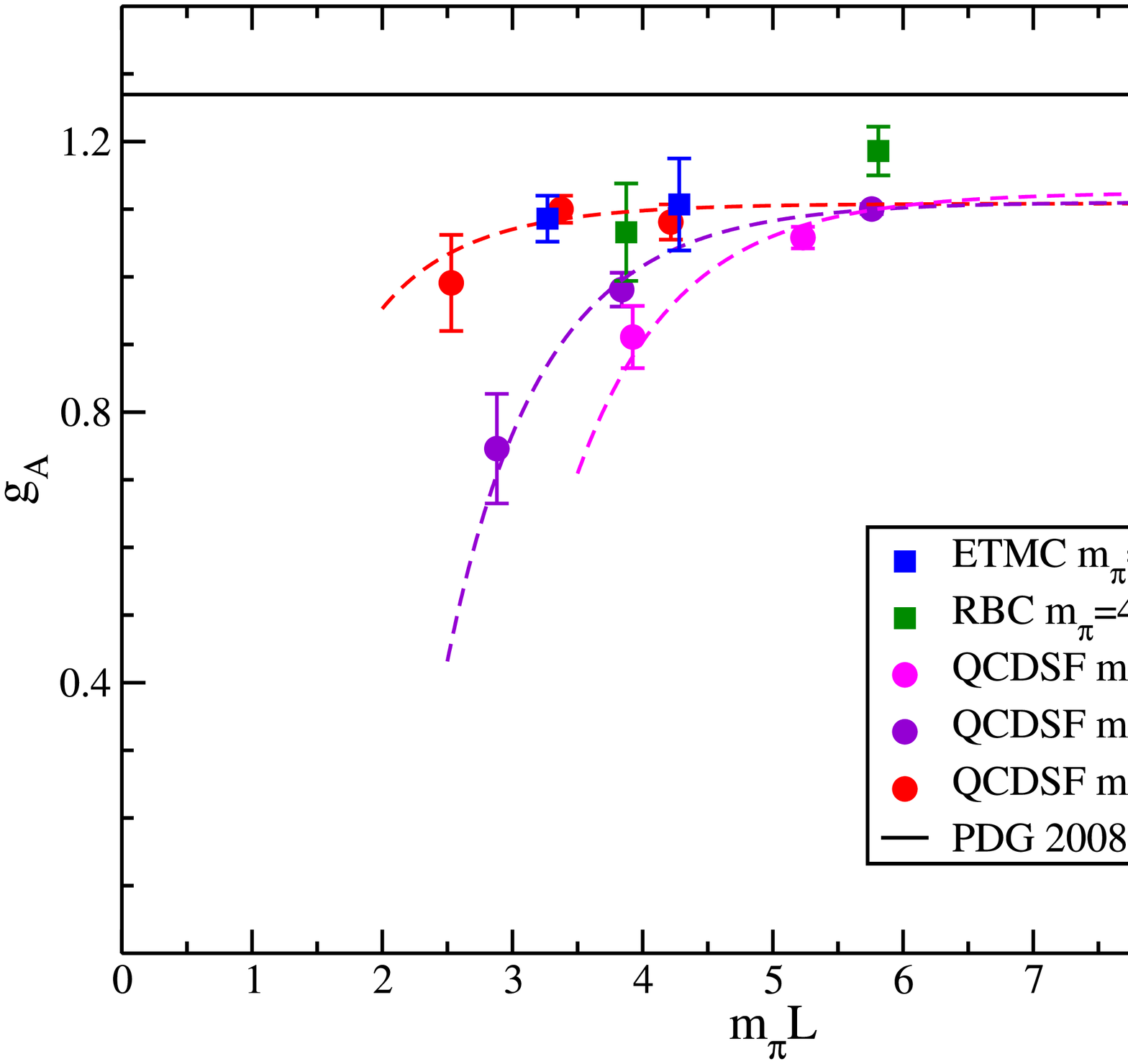}
\caption{Finite size effects in $\ga$.  The $m_\pi L$ dependence is
  shown for several calculations.  The results at the heaviest pion
  masses indicate a significant finite size effect in $\ga$,
  suggesting that $m_\pi L$ of $6$ may be necessary.  But the light
  $m_\pi$ results of QCDSF and ETMC may in fact be showing a weakening
  of the finite size effects in $\ga$ as measured in terms of $m_\pi
  L$.  The lattice results and experimental measurement are the same
  as in \Fig{\protect\ref{worldga}}, and the curves are simple scaling
  fits to guide the eye.}
\label{fsga}
\end{center}
\end{figure}

In \Fig{\ref{fsga}},
I examine several finite size studies that are available for $\ga$.
The results for the heaviest values of the pion mass show a strong
finite size effect.  This has led to the current view that $m_\pi L >
6$ may in fact be needed to reliably determine $\ga$.  However, we find
that, as quantified by $m_\pi L$, the finite size effects are
diminishing as $m_\pi$ is lowered.  This can be seen quite clearly in
\Fig{\ref{fsga}} by examining the lighter pion mass calculations of
both QCDSF and ETMC.  It is possible that the volumes for these
calculations at $m_\pi= 310~\mev$ or $260~\mev$ may still be too small
to see the asymptotic volume dependence.  We can also consult chiral
perturbation theory, which does allow for this sort of behavior;
however, given that $\ga$ is essentially flat for the largest volumes,
one might reasonably question the use of chiral perturbation theory at
these pion masses.

\begin{figure}[t]
\begin{center}
\includegraphics[width=\picwidthtwo,angle=\picangletwo]{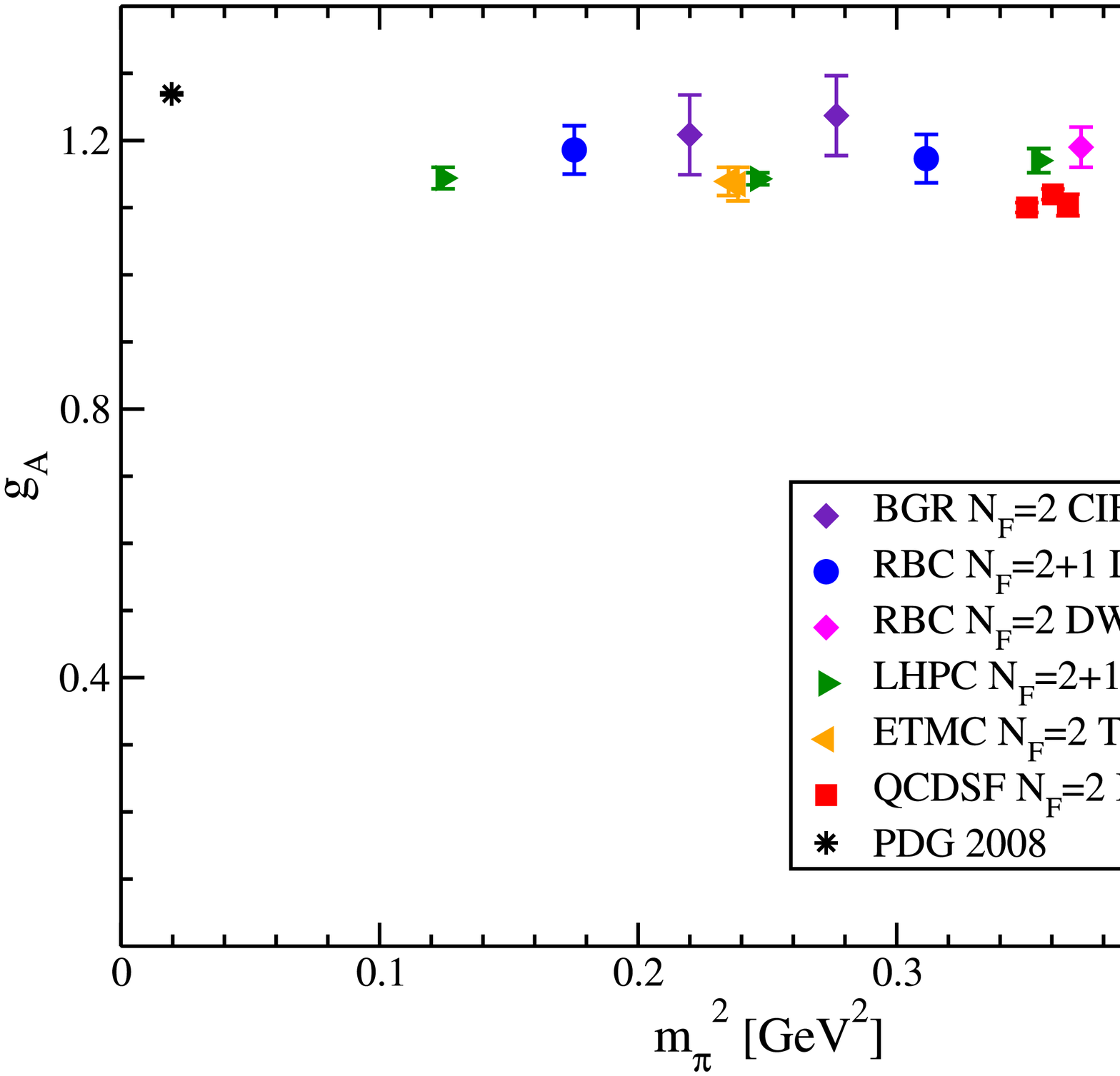}
\caption{Results for $g_A$ with $m_\pi L>6$.  The lattice results from
  \Fig{\protect\ref{worldga}} that satisfy $m_\pi L>6$ are shown.  All
  the lattice calculations, but one, are in statistical agreement.
  The results still underestimate the experimental value, but the
  discrepancy is much less severe than in $\x{u-d}$.  The QCDSF
  results are just slightly low.  This may indicate some small
  systematic error in the renormalization or it may simply illustrate
  the lattice artifacts that are present in all the calculations.}
\label{largevolga}
\end{center}
\end{figure}

In order to attempt to estimate the infinite volume limits for the
lattice calculations presented here, I simply require $m_\pi L > 6$.
This may turn out to be an excessive requirement for low $m_\pi$,
which would actually be good news regarding finite size effects, but
it is clearly required for heavier pion masses.  The lattice results
satisfying this are shown in \Fig{\ref{largevolga}}.
As is clear from the figure, this restriction is very severe, however,
the resulting lattice calculations show a strong level of agreement
amongst themselves.  The results are still lower than the experimental
measurement, but, unlike for $\x{u-d}$, only a mild curvature is
required to reconcile the current calculations with the physical
limit.  In particular, notice that, with one exception, there is no
systematic shift between the various collaborations.  As mentioned
earlier, the renormalization of $\ga$ is generically easier than that
of $\x{u-d}$, and this lends a bit more support to the hypothesis that
differences in renormalization may be driving part of the variation of
$\x{u-d}$ in \Fig{\ref{largevolx}}.  The one small exception is the
result of QCDSF which is just slightly low compared to all the other
calculations.  However, this may be consistent with a small discrepancy
in $f_\pi$, which renormalizes with the same factor as $\ga$, that is
also present for QCDSF~\cite{Jansen:2008vs}.

\section{Nucleon Charge}

The charge of the nucleon, or any hadron, again appears to be an
essentially trivial topic.  There is no doubt that the proton has one
unit of charge and that the neutron is, well, neutral.  The certainty
in our understanding of the nucleon's charge arises from the flavor
symmetries of QCD, however, the distribution of this charge in the
spatial degrees of freedom of the nucleon's constituents is not
prescribed by symmetry and provides another probe of the dynamics of
QCD within the nucleon.  This idea leads to a broad range of
calculations on the lattice and quickly involves the generalized
parton distributions, but to avoid excessive complications and to
explain the concepts in the simplest setting, I will discuss just the
neutron charge~\cite{Miller:2007uy}.

\subsection{Neutron Transverse Charge Distribution}

Much of the interest in form factors originates with the
interpretations that we can assign to them.  There is now a rich field
theoretic discussion dedicated to this issue, but in order to maintain
our footing and retain a strong connection to the experimental
measurements, it is valuable to remember that ultimately, the form
factors parametrize the QCD contributions to cross sections,
independent of any interpretation we attach to them.
\bd
\frac{d\sigma}{d\Omega} = \left(\frac{d\sigma}{d\Omega}\right)_\mathrm{Mott}
\left\{ F_1^2 + \tau F_2^2 + 2\tau(F_1+F_2)^2\tan^2(\theta/2) \right\}
\ed
Here $F_1(q^2)$ and $F_2(q^2)$ are the nucleon form factors and
$\tau=q^2/(4m^2)$.  For some in our community, this is where the
discussion ends.  That is, of course, a perfectly valid point of view.
However, the understanding provided to us by the generalized parton
distributions and the relationship to transverse quark
distributions~\cite{Burkardt:2000za} allows us to go further.

\begin{figure}[t]
\begin{center}
\includegraphics[width=\picwidthone,angle=\picangleone]{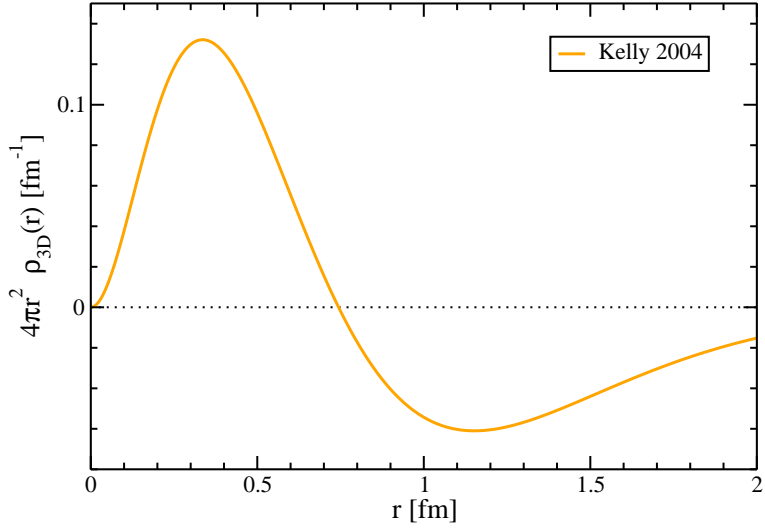}
\caption{Naive $3d$ neutron charge distribution.  The $3d$ Fourier
  transform of the $G_E$ form factor gives the rest frame charge
  distribution of the nucleon up to relativistic corrections.  The
  negative distribution at large $r$ is interpreted as the pion cloud
  arising from the $n\rightarrow p\pi^-$ fluctuations.  This gives
  rise to a negative charge radius for the neutron.  The curves were
  generated by numerically integrating the Kelly 2004 parametrization
  of the nucleon form factors~\cite{Kelly:2004hm}.}
\label{3dcharge}
\end{center}
\end{figure}

Historically the form factors were interpreted as charge and
magnetization densities.  In particular the charge density was
constructed from $G_E = F_1 - \tau F_2$ as
\bd
\rho_{3D}(\vec{r}) =
\int\!\! \frac{d^3\vec{q}}{(2\pi)^3}\,\,\, 
e^{-i\vec{r}\cdot\vec{q}}\, \frac{m}{\sqrt{m^2+\vec{q}^2}} G_E(\vec{q}^2)\,.
\ed
This is plotted for the neutron in \Fig{\ref{3dcharge}}.
There we see that the charge distribution is negative for large $r$.
This is usually understood in terms of the pion cloud generated by
$n\rightarrow p\pi^-$ fluctuations.  Overall neutrality of the neutron
forces the core to be positive leading to a negative charge radius.

\begin{figure}[t]
\begin{center}
\includegraphics[width=\picwidthone,angle=\picangleone]{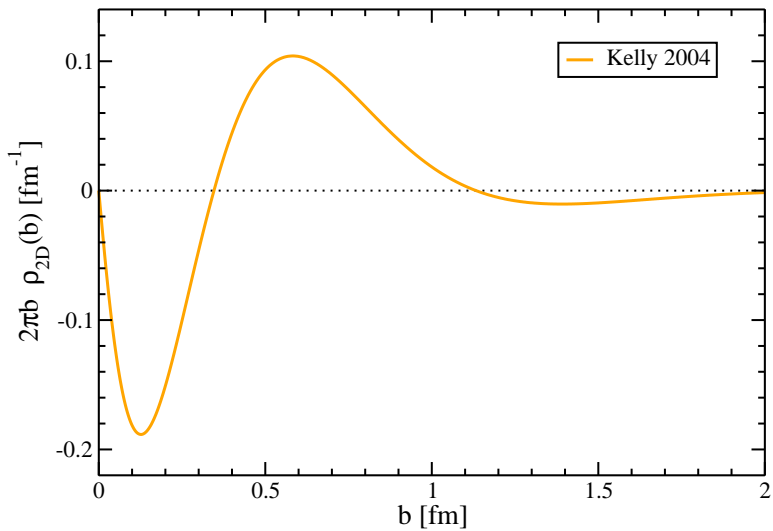}
\caption{Transverse 2d neutron charge distribution.  The $2d$ Fourier
  transform of the $F_1$ form factor results in the transverse charge
  distribution of the nucleon.  Relativistic corrections are absent
  but the distribution must be understood in the infinite momentum
  frame.  Notice the negative core, in contrast to
  \Fig{\protect\ref{3dcharge}}.  Also the charge radius is now
  positive in this frame.  The distribution was constructed from the
  same form factor parametrization in \Fig{\protect\ref{3dcharge}}.}
\label{2dcharge}
\end{center}
\end{figure}

But this picture is spoiled by relativistic corrections that enter as
$(m\langle r\rangle)^{-1}$, where $\langle r\rangle$ is some relevant
length scale for the charge distribution.  These corrections would
vanish in the non-relativistic limit $m\rightarrow\infty$ but are
substantial for the nucleon.  The introduction of the transverse
distributions in the infinite-momentum frame (IMF) eliminates these
corrections, which enter now as $(p_z\langle r\rangle)^{-1}$ with
$p_z\rightarrow\infty$ defining the IMF.  The transverse
distributions~\cite{Burkardt:2000za} are given by
\bd
\rho_{2D}(\vec{b}) =
\int\! \frac{d^2\vec{q}_\perp}{(2\pi)^2}\,\,\, 
e^{-i\vec{b}\cdot\vec{q}_\perp}\, F_1(\vec{q}_\perp^2)\,.
\ed
This is shown for the neutron in \Fig{\ref{2dcharge}}.
The distribution still has a negative tail for large $r$, but now we
find a negative core.  Furthermore, the charge radius is positive for
the transverse charge distribution.  Understanding this new view of
the neutron is a challenge onto itself.  Additionally, the
introduction of the transverse coordinates in a field-theoretically
clean manner opens many new avenues for investigation of hadron
structure from lattice QCD.

\subsection{Lattice Calculation of $\mathbf{\rsq{u-d}}$}

As for the nucleon momentum and spin structure, I again introduce a
benchmark observable that allows us to summarize the status of lattice
calculations of form factors.  Here I choose the isovector charge
radius, $\langle r^2 \rangle_{u-d}$.  As before, the flavor structure
is $u-d$, so disconnected diagrams again vanish.  Furthermore, this
observable enters matrix elements of the vector current.  These two
facts combine to make $\langle r^2 \rangle_{u-d}$ an accurate quantity
to calculate on the lattice.  Additionally, all the lattice
calculations can now use an exactly conserved vector current.  This
eliminates the need for any renormalization and simplifies the
calculation significantly.

The form of the nucleon's electromagnetic interaction is given by
\bd
\langle p^\prime, s^\prime | J^\mu | p, s \rangle =
\overline{u}(p^\prime,s^\prime) \left\{ \gamma^\mu F_1(q^2)
+ i\sigma^{\mu\nu}\frac{q_\nu}{2m} F_2(q^2) \right\} u(p,s)
\ed
and the charge radius is then given by
\bd
\langle r^2\rangle_{u-d} = -6 \left.\frac{dF^{u-d}_1}{dq^2}\right|_{q^2=0}\,.
\ed
The factor of $6$ is the conventional choice, but a factor of $4$
would instead give the correct transverse charge radius in the IMF.
To avoid confusion, we stick to the conventional choice, but we should
keep in mind the theoretically clean interpretation that is available
for the transverse distribution.

\begin{figure}[t]
\begin{center}
\includegraphics[width=\picwidthtwo,angle=\picangletwo]{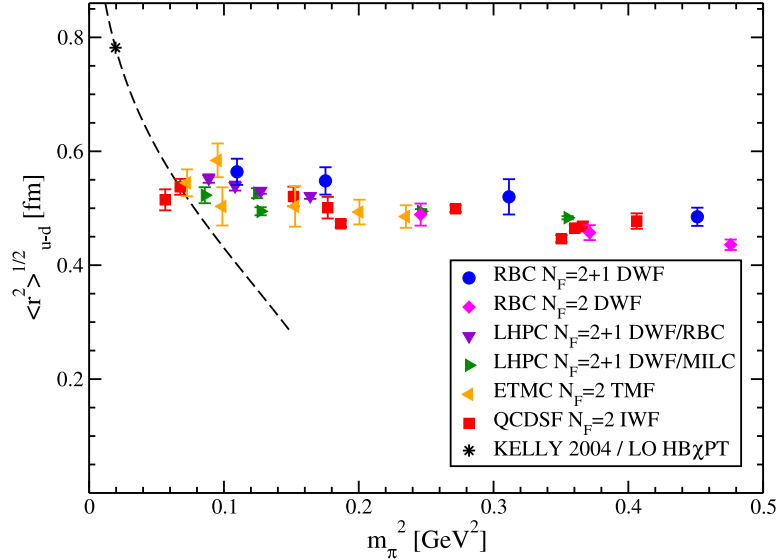}
\caption{World's dynamical results for $\langle r^2
  \rangle_{u-d}^{1/2}$.  The lattice calculations are
  \cite{Yamazaki:2009zq} (RBC $N_\mathrm{F}=2+1$),
  \cite{Lin:2008uz,Aoki:2004ht} (RBC $N_\mathrm{F}=2$), \cite{:2010jn}
  (LHPC/MILC), \cite{Syritsyn:2009mx} (LHPC/RBC), \cite{korzec:email}
  (ETMC) and \cite{james:email} (QCDSF).  The experimental result
  comes from the Kelly 2004 parametrization of the form
  factors~\cite{Kelly:2004hm}.  The leading order heavy baryon chiral
  perturbation theory prediction (LO HB$\chi$PT) contains no unknown
  low energy constants other than $\langle r^2 \rangle_{u-d}^{1/2}$
  and is logarithmically divergent as $m_\pi\rightarrow 0$.  One
  might, therefore, expect substantial finite size effects in this
  quantity, but this does not appear to be the case.}
\label{worldrsq}
\end{center}
\end{figure}

In \Fig{\ref{worldrsq}}
I gather all the full QCD results for $\rsq{u-d}$.  It is immediately
evident that all the lattice calculations show a remarkable agreement
for this observable.  The strong scatter that is present for $\x{u-d}$
and to a lesser extent $\ga$ is clearly absent for $\rsq{u-d}$.  This
is particularly surprising given the fact that this quantity is
expected to diverge in the chiral limit as indicated by the curve in
\Fig{\ref{worldrsq}}.  We might even expect large finite size effects
in such a quantity.  However, we must keep in mind that $\rsq{u-d}$ is
calculated from the slope of $F_1(q^2)$ at $q^2=0$ and that the
momentum quantization imposed by a finite volume forces us to extract
this slope using the first non-zero value of $q^2$, which satisfies $q
L\approx 2\pi$.  We may, in fact, be extrapolating from a region with
weaker $L$ dependence through the region with stronger $L$ dependence.
Ultimately, whatever the explanation, we find a near absence of volume
dependence for $\rsq{u-d}$.

\begin{figure}[t]
\begin{center}
\includegraphics[width=\picwidthtwo,angle=\picangletwo]{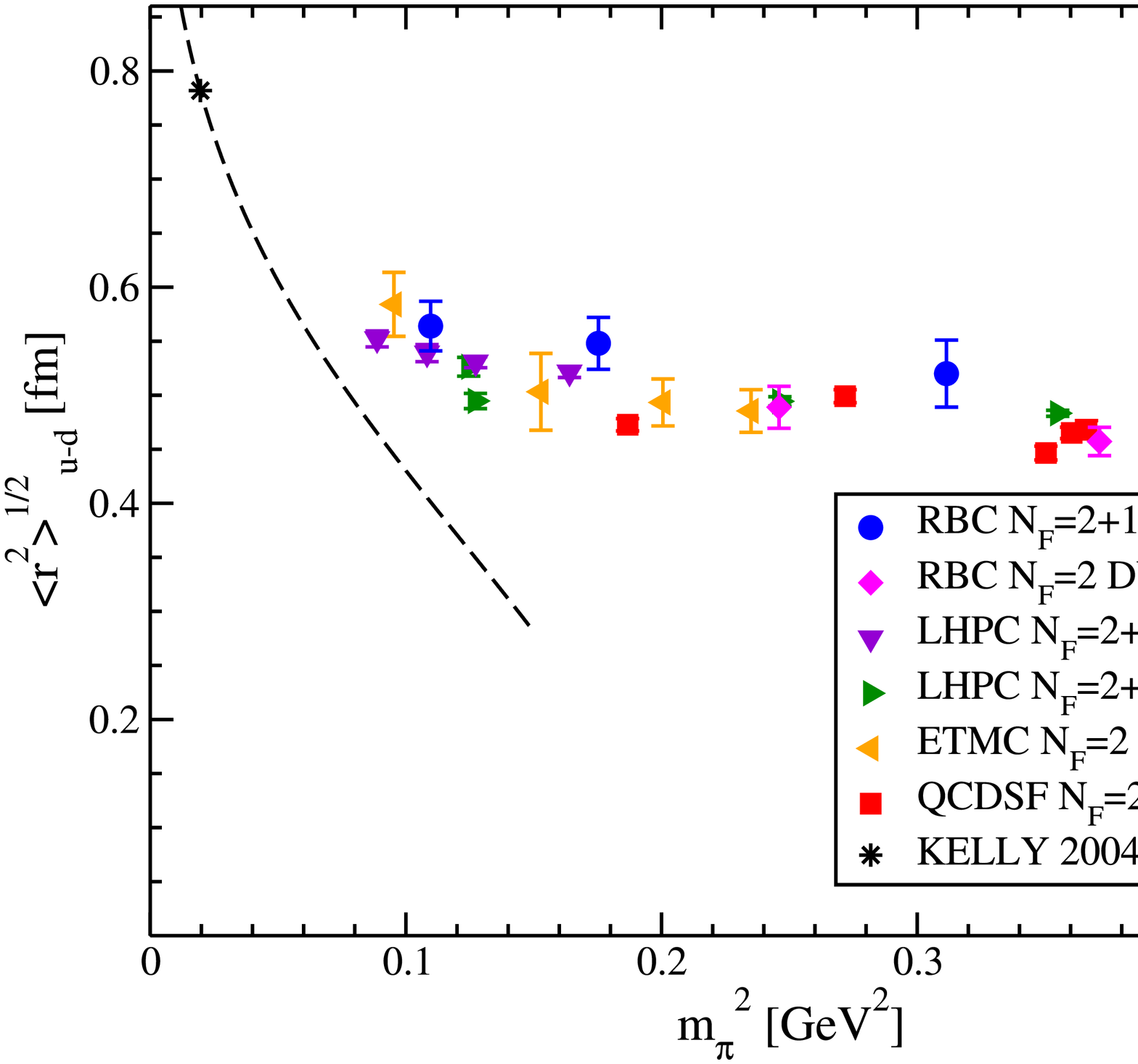}
\caption{Large volume results for $\langle r^2 \rangle_{u-d}^{1/2}$.
  Due to momentum quantization at finite volume, one can not easily
  calculate the form factors at the low $Q^2$ needed to determine
  $\langle r^2 \rangle_{u-d}^{1/2}$ while also working at small
  volume.  Lacking any detailed knowledge of the finite size effects
  in $\langle r^2 \rangle_{u-d}^{1/2}$, only the results from
  \Fig{\protect\ref{worldrsq}} with $m_\pi L>4$ are shown.  There is
  general agreement among all the lattice groups and a mild increase
  as $m_\pi$ is lowered.  However $\langle r^2 \rangle_{u-d}^{1/2}$ is
  not yet rising nearly as fast as it needs to match onto the
  experimental measurement.}
\label{largevolr}
\end{center}
\end{figure}

Lacking any real guidance on the necessary volumes for $\rsq{u-d}$, I
simply follow the conventional wisdom and take $m_\pi L>4$.  These
lattice results are shown in \Fig{\ref{largevolr}}.
An optimist may find a mild increase, but this would be hard to defend
strongly.  Additionally, this observable requires much more than a
mild increase to agree with the experimental value at the physical
point.  But it does appear that all the calculations seem to agree.
Given that $\rsq{u-d}$ requires no renormalization, this is consistent
with the speculation suggesting that renormalization may be partially
culpable for the systematic variation in $\x{u-d}$ discussed earlier.
Despite the fact that this observable still shows a significant
discrepancy with the experimental result, the agreement amongst all
the lattice groups is encouraging.

\section{Conclusions and Outlook}

In order to illustrate the status and promise of lattice calculations
of hadron structure, I chose to review three benchmark observables
that represent the broad range of hadronic physics that we can address
on the lattice:\ the momentum fraction, axial charge and charge radius
of the nucleon.  Nearly all the lattice groups have pushed the pion
mass towards $300~\mev$ or lower and some are now breaking through
$200~\mev$.  This is an enormous accomplishment, but unfortunately
most hadronic observables are failing to show the long-sought chiral
curvature.  Independent of the expectations of chiral perturbation
theory, the three benchmark observables still do not show a convincing
approach to the experimental results.

The obvious conclusion is that yet lighter, and maybe even physical,
pion masses will be required.  This may be the case, but we should not
do this at the expense of other systematic errors.  Finite size
effects must remain a concern as we lower $m_\pi$.  Both $\x{u-d}$ and
$\ga$ demonstrate that the volume dependence can have a non-trivial
dependence on $m_\pi$ and our conventional rule-of-thumb that $m_\pi
L\approx4$ may not necessarily be sufficient for all quantities.

Lattice artifacts are still a concern, if only because so few groups
have seriously examined them.  The agreement of the groups for
$\rsq{u-d}$ might indicate weak artifacts for some quantities, but the
systematic shifts between the groups for $\x{u-d}$ is a warning that
renormalization, an essential part of the continuum limit for most
quantities of interest, may be partially responsible for these
problems.  The role of renormalization is further implicated by the
general agreement of the groups for the renormalization free ratio of
$\x{u-d}/\x{u-d}^\mathrm{ref}$.  In fact, the progression from
$\x{u-d}$ (requiring a scale-dependent renormalization) to $\ga$
(requiring only a finite renormalization) to $\rsq{u-d}$ (requiring no
renormalization) will be a strong test of our control of
renormalization issues.

These systematic effects will be a part of any critical review of
lattice calculations, but we should remember the rich QCD physics that
provides the motivation for these calculations.  Each of the benchmark
observables is a stand-in for a broad range of experimentally relevant
and intellectually interesting observables:\ the decomposition of
momentum among the nucleon's constituents, the nature of the nucleon's
spin and the distribution of the nucleon's charge.  Apart from these
observables, Lattice 2009 saw a sizeable effort to study strangeness
in the nucleon, results for the pion, delta and other hadrons, and a
range of exploratory calculations.  Clean control of systematics and a
continued push toward the physical pion mass will ultimately allow
detailed calculations that will complement the experimental programs
and advance our insight into hadron structure.

\acknowledgments

I gratefully appreciate the many conversations and email exchanges
with colleagues while preparing for the Lattice 2009 conference.  I
would like to thank D.~Alexandrou, G.~Bali, W.~Bietenholz, T.~Blum,
T.~Doi, M.~Engelhardt, W.~Freeman, Ph.~H\"agler, P.-A.~Harraud,
T.~Hemmert, K.~Jansen, T.~Kaneko, T.~Korzec, M.~Lin, Z.~Liu,
D.~Mohler, B.~Musch, J.~Negele, H.~Ohki, S.~Ohta, J.~Osborn,
G.~Schierholz, W.~Schroers, T.~Streuer, K.~Takeda, N.~Ukita,
A.~Walker-Loud and J.~Zanotti.  Additionally, this work was supported
by the DFG Sonderforschungsbereich/Transregio SFB/TR9-03

\bibliographystyle{unsrt}
\bibliography{lat09.bib}

\end{document}